# A Hybrid Wired/Wireless Deterministic Network for Smart Grid

Bin Hu and Hamid Gharavi
Advanced Network Technologies
National Institute of Standards and Technology
Gaithersburg, USA
Emails: [bhu, Gharavi]@nist.gov

*Abstract*—**With the rapid growth of time-critical applications in smart grid, robotics, autonomous vehicles, and industrial automation, demand for high reliability, low latency and strictly bounded jitter is sharply increasing. High-precision time synchronization communications, such as Time Triggered Ethernet (TTE), have been successfully developed for wired networks. However, the high cost of deploying additional equipment and extra wiring limits the scalability of these networks. Therefore, in this paper, a hybrid wired/wireless high-precision time synchronization network based on a combination of high-speed TTE and 5G Ultra-Reliable and Low-Latency Communications (URLLC) is proposed. The main motivation is to comply with the low latency, low jitter, and high reliability requirements of time critical applications, such as smart grid synchrophasor communications. Therefore, in the proposed hybrid network architecture, a high-speed TTE is considered as the main bus (i.e., backbone network), whereas a Precision Time Protocol (PTP) aided 5G-URLLC-based wireless access is used as a sub-network. The main challenge is to achieve interoperability between the PTP aided URLLC and the TTE, while ensuring high precision timing and synchronization. The simulation results demonstrate the impact of the PTP-aided URLLC in maintaining network reliability, latency, and jitter in full coordination with the TTE-network.**

## I. Introduction

The fourth industrial revolution, Industry 4.0 [1], is experiencing high speed development. It uses sophisticated information and communication technologies for intelligent networking of machines and processors. In Industry 4.0, time-critical applications execute tasks with hard real-time constraints, where given deadlines must be satisfied along with strict constraints on reliability, latency, and jitter. One of the key attributes of Industry 4.0 is the power grid, which is rapidly evolving to a smart grid. For instance, the next generation of smart grid networking technologies need to provide highly reliable and low latency synchrophasor communications [2]. Phasor Measurement Units (PMUs) are deployed in a large number to monitor the health of a power grid in real time, hence requiring a secure and reliable communication infrastructure. With the proliferation of Ethernet-based networks in substation automation systems, timing protocols, such as Network Time Protocol (NTP), IRIG-B and one pulse per second (1-PPS), may not be sufficient to achieve timing accuracy < 1 μs as well as performance reliability > 99.99 %, and latency < 10 ms [3]. At the same time, improving the quality of synchrophasor measurements, especially for a substation automation system (SAS), would require a secure and deterministic communication infrastructure with much greater timing precision. In particular, with widening PMU installations in massive numbers for wide area monitoring, demands for higher communication bandwidth continues to grow. To comply with such stringent requirements, we consider a deterministic time-sensitive network (TSN), namely Time-triggered Ethernet (TTE) capable of achieving constant latency and minimum jitter. This is mainly because TTE compensates transmission jitter that is introduced during traffic propagation and operation in the switches [4]. Furthermore, to remotely control sensor devices, such as Intelligent Electronics Devices (IEDs), it is important to utilize a wireless technology with high reliability and low delay. Therefore, the TTE based network architecture should be able to provide a wireless solution for interconnecting critical and noncritical sensory data wherever the wired access is not available. Under these conditions, it is important to utilize a wireless technology that can guarantee reliability, accuracy, and low delay information exchanges. Therefore, an integration of wired and wireless networks that can collectively coordinate timing and synchronization with the TTE backbone network is a key technical challenge.

Currently there are limited wireless options available that can match the high-precision timing and reliability of the TTE network. For instance, there is an ongoing standards activity to offer a new generation of wireless local area networks (WLAN), referred to as WiFi-7 [5]. The future WiFi-7 will be based on the IEEE 802.11be amendment, which will build upon 802.11ax, focusing on WLAN indoor and outdoor operation with stationary and pedestrian speeds. More importantly, it aims at providing time-sensitive capabilities to enable high reliability and low latency in the 2.4, 5, and 6 GHz license-exempt spectrum bands [5]. The IEEE 802.11be PHY and MAC functionalities will be used for low latency operation. A timing measurement procedure, similar to PTP, can be used to achieve high precision clock synchronization between a primary node (i.e., the access point) and its secondary nodes (i.e., the stations). While WiFi-7 is still in its infancy, 5G URLLC [6-8] will provide the most reliable and widely available wireless communications. Bear in mind that among the three application scenarios covered by 5G, we consider the URLLC for our wireless expansion. Obviously, this would require designing a highly synchronized air interface where TTE, as the main bus, can provide wireless access via URLLC. URLLC [6-8] is designed to support the stringent latency and reliability requirements of mission-critical applications. This includes providing time synchronization across the network, Orthogonal Frequency Division Multiple Access (OFDMA) reservation to reduce latency, scheduled access for better control of channel access, and link adaptation. Such important features make 5G URLLC highly suitable for integration with TTE for time-sensitive communications. TTE is based on the AS6802 standard [9], which has been developed for integrated systems and safety-related applications. However, implementing AS6802-based high precision time synchronization in URLLC requires the adaptation and installation of TTE controllers in all components of the network.

For instance, PTP [10] can be used to enable high precision time synchronization in 5G URLLC wireless networks. PTP is a high precision synchronization protocol and is applicable to networks using multicast messaging. It has been developed to synchronize independent end devices running on distributed network nodes to a primary clock using hardware timestamping to minimize delay variability in the network stacks of the devices. PTP messages that carry precise timing information are exchanged between a reference time source through a network with full PTP support to an end node to provide sub-microsecond precision. Its precision depends mainly on the accuracy of the timestamps used for synchronization [10].

The challenge, however, is to how to establish over the air high precision timing and synchronization between the TTE and the URLLC. To achieve scalable integration of both protocols, we present a hybrid wired/wireless deterministic network where timing synchronization, resource reservation and scheduling is used to guarantee extremely low data loss rates, low jitter, and bounded latency.

We then demonstrate the effectiveness of the proposed approach for smart grid synchrophasor communications where precise timing and synchronization with high reliability would be essential.

After a brief overview of the PTP protocol and 5G URLLC, we present the air interface aspect of our proposed hybrid Time-Sensitive network (TSN) in Section II. The hybrid network architecture is then introduced in Section III. To assess the performance of the proposed network, especially for smart grid synchrophasor applications in terms of latency, jitter, and service reliability, we present the simulation results in Section IV. This is then followed by the conclusion in Section V.

## II. PTP AIDED 5G URLLC WIRELESS NETWORK

High precision time synchronization technology has been widely used to provide time critical services in smart grid, industrial automation, precision measurement, and control [1]. PTP is defined by the IEEE 1588v2 standard in order to provide sub-microsecond network time synchronization with high accuracy, which makes it suitable for time sensitive critical applications.

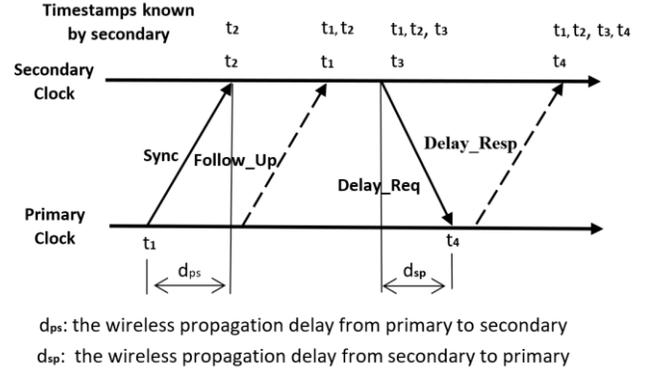

Fig. 1: PTP synchronization message exchange.

PTP messages, using the end-to-end delay measurement mechanism, exchanged between a primary and a secondary clock for synchronization are depicted in Fig. 1. The PTP primary clock sends a synchronization (SYNC) message once every synchronization interval. The SYNC message contains the clock identification information and the specified accuracy of the primary clock based on the current reference source, as well as an estimated timestamp, $t_1$, of the SYNC message transmission time. Since the estimated timestamp, $t_1$, in the SYNC message may not be accurate enough, the primary clock will send a follow-up (Follow_Up) message with a more precise value of the timestamp $t_1$ to the secondary clock. As soon as the SYNC message is received, the secondary clock will timestamp the reception time, $t_2$, using its own clock. The difference between the SYNC message's departure timestamp, $t_1$, and arrival timestamp, $t_2$, is a combination of the secondary clock's offset from the primary clock and the wireless channel propagation delay. The secondary clock then sends back a delay request (Delay_Req) message to the primary clock and stores the departure time with timestamp, $t_3$. When a primary clock receives the Delay-Req message, it sends a delay response (Delay-Resp) message to the secondary clock, which contains the timestamp, $t_4$, of the reception time of the corresponding Delay_Req message.

At this point, the secondary clock obtains four timestamps, which can then be used to derive the timing offset between the two clocks. Under these conditions, the wireless channel propagation delay can be derived as:

$$\text{propagation delay} = \frac{(t_4-t_3)+(t_2-t_1)}{2}. \quad (1)$$

The computation of this propagation delay is based on the assumption that the primary-to-secondary and secondary-to-primary propagation delays are equal. User Equipment (UE) with high mobility will experience an asymmetric wireless link, which is caused by the rapidly changing distance. However, in URLLC the end-to-end delay between a base station (i.e., Next-Generation Node B (gNB)) and UEs is bounded within 1 ms, the distance changed in such short a period is around 0.05 m for a fast-moving UE at the speed of 50 m/s. The delay asymmetry is less than $1 \times e^{-10}$ s and is negligible. Generally, there is limited delay asymmetry in the employed URLLC wireless network.

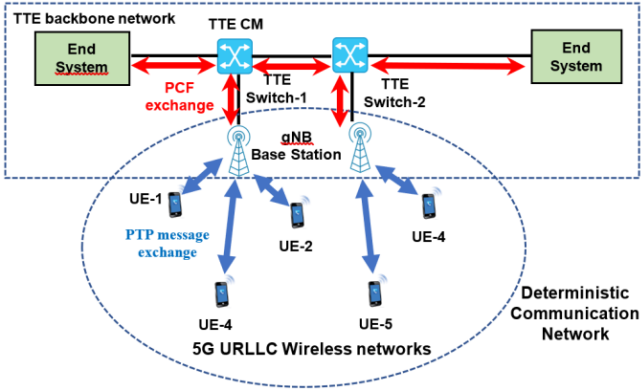

Fig. 2. Hybrid wired/wireless deterministic communication network for time critical applications.

Under the above conditions, the clock offset between the primary clock and the secondary clock is then expressed as:

$$\text{Offset} = t_2 - t_1 - \text{ propagation delay}. \quad (2)$$

By exchanging these synchronization messages, the end devices would be able to accurately measure offset and then update their clocks accordingly in order to achieve synchronization with the primary clock.

PTP, which was originally proposed for wired networks, has been successfully used for time critical wireless applications, such as URLLC [11, 12]. It should be noted that URLLC is served with large numerologies where short time slots and larger Configurable Subcarrier Spacing (SCS) are used to meet strict latency requirements. Smart grid and industrial automation are two important use case scenarios of URLLC [3]. In both scenarios high reliability, low latency, and strictly bounded jitter would be essential as time synchronization is intrinsic to real-time coordination and interaction among devices.

To obtain high precision synchronization between a base station (i.e., gNB) and its associated UEs (e.g., PMUs or/and other sensory devices), we employ PTP in 5G URLLC wireless networks in conjunction with the TTE network. In this case, a gNB is used as the primary clock to provide time reference inside the URLLC domain, while each user equipment (UE) acts as a secondary clock. As show in Fig. 2, a gNB primary acquires a reference time from an external clock, which is the TTE compression master in our proposed model (see Section III). A UE acquires the reference time from the gNB primary based on the PTP and the selected path delay measurement mechanism. Specifically, the gNB base station as a PTP primary sends SYNC and Follow-Up messages to UEs (i.e., PTP secondaries). A UE, as a PTP secondary, sends back a Delay-Req message to the gNB primary. This is then followed by receiving a Delay-Resp message as shown in Fig. 1.

We should point out that in the above message exchanges dedicated Radio Resource Control (RRC) signaling is used to establish a primary-secondary relationship between a gNB and its associated UEs. Bear in mind that RRC signaling in URLLC is capable of transmitting small packets, such as the synchronization messages, with ultra-low latency and high reliability. The main function of RRC signaling includes connection establishment, maintenance, and release. By using the Dedicated Control Channel (DCCH), RRC signaling can prevent collisions and achieve high precision clock synchronization [13]. After establishing an RRC-based connection, the gNB primary clock periodically unicasts SYNC, Follow-Up, and Delay-Resp messages to the UE secondary clock via its downlink dedicated control channel (i.e., DL-DCCH). Subsequently, the UE sends its Delay-Req message back to the gNB primary clock through the uplink dedicated control channel (i.e., UL-DCCH). Furthermore, by classifying these messages as time critical, they will be handled with the highest priority.

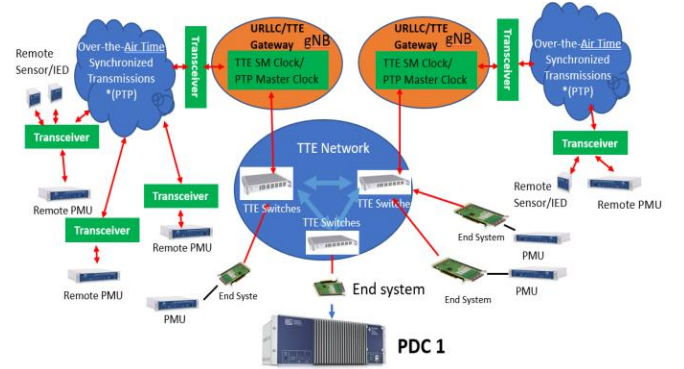

Fig. 3. Proposed hybrid deterministic network for smart grid.

To ensure high-precision time synchronization, URLLC uses the OFDMA reservation to reduce latency, scheduled access, and bandwidth reservation for better control of channel access and link adaptation [6]. Furthermore, link delay measurements in PTP can be used to measure peer-to-peer delay, where PDelay_Req and PDelay_Resp are used. As a result, time critical messages, such as real-time synchrophasor data in smart grid, can be sent and received at pre-configured time instances where their transmission reliability, latency, and jitter can be guaranteed. These capabilities make the PTP-aided URLLC highly suitable for integration with TTE to enable deterministic communications.

### III. PROPOSED HYBRID DETERMINISTIC NETWORK

The proposed hybrid network is shown in Fig. 3. It consists of TTE switches, gNB stations, and end-systems. Each gNB interconnects a PTP aided URLLC wireless sub-network in coordination with the TTE backbone network. UEs (i.e., PMUs) are distributed throughout the network where their data are transmitted to the Phasor Data Collector (PDC) via the hybrid wired/wireless network. TTE, due to its important features of low latency, guaranteed reliability, and high bandwidths, is used as the backbone network as shown in Fig.3. To achieve high precision time synchronization, TTE protocol uses dedicated messages called protocol control frames (PCF) to establish and maintain system-wide clock synchronization amongst all the nodes in the TTE backbone network. The Synchronization entities consist of Compression Master (CM) and Synchronization Master (SM) or synchronization clients (SC), which are selected based on the system architecture. Switches are normally configured as CM and end-systems as SM. If neither is configured as the synchronization master or compression master, they will function as a synchronization client. The precision of the synchronization depends on the clock drift of each local clock at synchronization masters (e.g.

100 parts per million (ppm)) and the integration cycle [14]. Smaller integration cycles result in smaller offset, hence higher precision synchronization. In the proposed architecture, the TTE integration cycle is set to the same value as the PTP synchronization interval to avoid any inconsistencies.

As shown in Fig. 2, in the TTE domain, a TTE switch is configured as a CM, while its associated gNB and other end-systems are configured as SMs. The gNB acquires the global time from the TTE switch by periodically exchanging PCFs in the TTE backbone domain. Each gNB acts as the PTP primary clock in its URLLC domain, while the associated UEs function as the PTP secondary clocks. After synchronizing with the TTE switch, the gNB primary exchanges PTP messages with its associated UE secondaries to distribute the global time. In this way, all the UEs, gNB stations, and other end-systems in the proposed hybrid network will be fully synchronized to the global time generated by the TTE switch.

In this network, each gNB station acts as a gateway between the TTE backbone network and the URLLC wireless subnetwork. In the case of smart grid communications, all PMUs are synchronized and generate measurement data at the same time and with high precision. These measurements are transmitted at pre-configured times to the associated gNB station using OFDMA/bandwidth reservation to avoid any delay caused by interference and collision. Since the link delay between a gNB station and its associated PMUs can be estimated using the link delay measurement procedure of the PTP, gNB stations can receive synchrophasors data at pre-configured times with high reliability and in a precise manner. The synchrophasor data are classified as Time-triggered messages by gNB, where the EtherType field of the Ethernet frame is set as 0x88d7. Each gNB sends these TT messages to a remote Phasor Data Concentrator (PDC) (i.e., one of the end-systems) at pre-configured times across the TTE backbone network. The propagation delay of the backbone network can then be measured precisely (e.g. 100 ns per link). In this way, the synchrophasors data can be transmitted in real-time and received with extremely high reliability, low latency, and strictly bound jitter.

## IV. SIMULATION

In this section, the proposed hybrid deterministic communication network shown in Fig. 3 was investigated for smart grid synchrophasor applications. Our simulation was carried out by using OMNET++ simulator and we developed PTP-aided URLLC to simulate the sub-network. In this example, the PDC was connected to the TTE switch as an end-system. Synchrophasor data, which were generated at 60 frames per second, were transmitted in real-time via the User Datagram Protocol (UDP) at the transport layer. URLLC packets each carrying about a 100 byte synchrophasor data frame were transmitted via uplinks to their local gNB station. After being converted to TT messages at the gNB station, these packets were sent to the destination PDC via TTE switches. We should point out that in this paper we are mainly concerned with providing over the air synchronization once the link is established between the gNB and TT-E gateway. This would require highly accurate synchronization with the TTE backbone network.

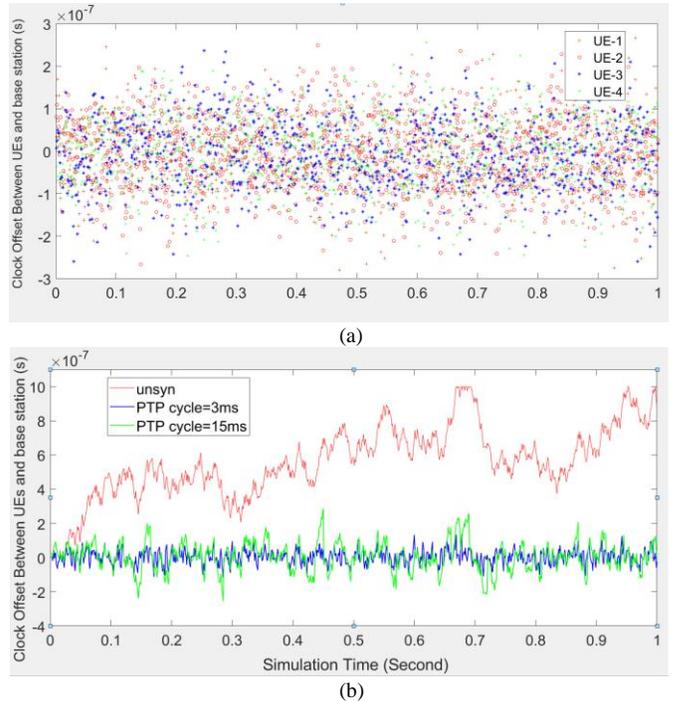

Fig. 4. Clock offset between UEs and the associated gNB station in a URLLC wireless sub-network, which is measured in every 1ms. The clock drift of all components is random between 50 ppm. (a): Clock offset measured in a PTP aided URLLC, where the synchronization interval is 3 ms. (b): Traditional URLLC vs PTP aided URLLC with synchronization interval being 3 ms and 15 ms.

In our simulation, the 3GPP channel model [15] was employed where different scenarios, such as rural, urban, indoor office, and indoor shopping malls can be selected. For simplicity, the rural scenario without shadowing and with the line of sight option was used.

All TTE links are bidirectional 100 Mbps links. The total bandwidth in each URLLC is 200 MHz, where the bandwidth part (BWP) is configured with numerology 4. Numerology is used by 5G to provide flexibility in the framing structure. Depending on the configured numerology, each subframe with a duration of 1 ms, is split into a variable number of slots. Ranging from 0 to 4. the main objective of numerology is to achieve a trade-off between latency and throughput under different types of traffic. URLLC uses numerology 4 with the shortest slot length in order to guarantee low latency and low jitter.

The drifts of TTE clocks and PMU clocks were configured as 50 or 100 parts per million (ppm). The propagation delay of the TTE backbone network was set at 100 ns per link. In a TTE network, virtual links are used as logical connections to route traffic from a sender to one or more receivers, which are pre-configured during the scheduling process. TTE switches use pre-defined forwarding tables to concatenate virtual links and create tree structures with one sender as the root and multiple receivers as the leaf nodes [7].

In Fig. 4, we evaluated the performance of the PTP aided URLLC and compare the results with traditional URLLC without PTP. The clock drift of all components was randomly distributed within 50 ppm. The clock offset between UEs (e.g., PMUs} and their associated gNB was measured every 1 ms. Fig. 4 (a) displays the clock offset measured in a PTP aided URLLC, where the synchronization interval was 3 ms. Using PTP, the jitter of all UEs was bounded within 600 ns. Fig. 4 (b) compares traditional URLLC with PTP aided URLLC, where the synchronization interval is either 3 ms or 15 ms. Using PTP significantly improved the jitter performance. The jitter caused by the clock offset is around 600 ns when the PTP synchronization interval is 15 ms. Next, we can decrease the PTP cycle time to reduce any offset between secondary clocks of UEs and the primary clock in the associated gNB station. When PTP synchronization interval is 3 ms, the jitter is around 200 ns. Smaller synchronization intervals lead to smaller offset and hence higher precision time synchronization.

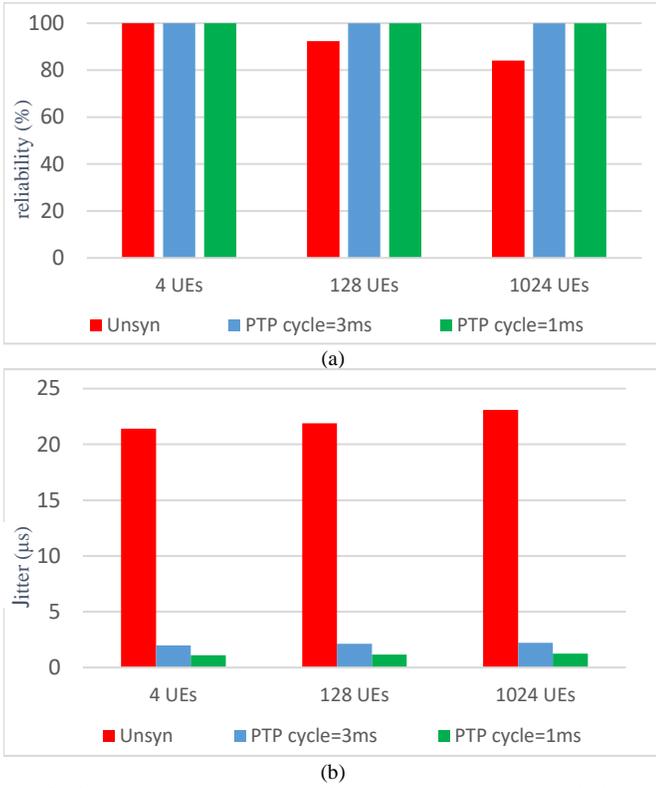

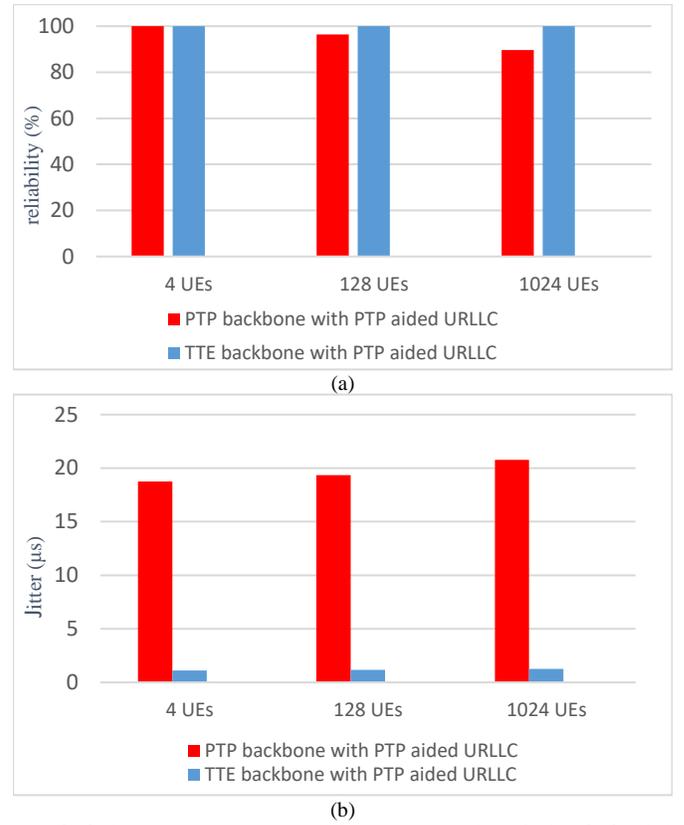

Fig. 5. Performance comparison of the TTE network with the PTP aided URLLC to that with traditional URLLC (a): reliability and (b): jitter performance. The clock drift is random between 50 ppm and the PTP synchronization interval is 1 ms or 3 ms.

In Fig. 5, we evaluated the performance of the proposed hybrid wired/wireless network and compare the results with the TTE backbone network when interfaced with a traditional URLLC sub-network. Under these conditions, the clock drift randomly fluctuates within 50 ppm. In these experiments, the TTE integration cycle is set to 1 ms or 3 ms in accordance with the PTP synchronization interval. As shown in Fig. 5, the proposed hybrid deterministic network indicated an extremely reliable performance (reliability > 99.99%). Its jitter is around 2 µs when the TTE integration cycle is 3 ms, or 1 µs when the TTE integration cycle is 1 ms. Conversely, the TTE backbone network with a traditional URLLC sub-network fails to provide deterministic communication due to an inherent lack of synchronization. As a result, the jitter increases by more than 20 µs and the reception time of the synchrophasor measurements cannot be accurately estimated. In addition, transmission reliability degrades when more PMUs (UEs) participate in sending synchrophasor data to the PDC.

Fig. 6. Performance comparison of the TTE network with the PTP aided URLLC to a PTP Ethernet network with PTP aided URLLC (a): reliability and (b): jitter performance. The clock drift is random between 50 ppm and the PTP synchronization interval is 1 ms.

Finally, we compare the performance of the proposed hybrid network with a PTP-assisted standard Ethernet network with PTP aided URLLC. In the case of the proposed TTE-based hybrid network, the TTE integration cycle and the PTP synchronization interval were set to 1 ms. The clock drift randomly fluctuated within 50 ppm. In the case of the PTP-assisted standard Ethernet network with PTP aided URLLC, two Ethernet switches, two gNB stations, and two other Ethernet clients composed a PTP backbone network. Each gNB station interconnects a PTP aided URLLC wireless sub-network with the PTP backbone network. PMUs were distributed in the URLLC wireless network as UEs. A PDC, which receives the PMUs data, is connected to the Ethernet switch as an Ethernet client. One of the Ethernet switches was selected as the primary clock. All PMUs, gNB stations, and the PDC are synchronized to the primary clock by exchanging PTP synchronization messages.

Both simulated network architectures, with full on-path PTP timing support, achieved high precision time synchronization. Their clock offset were around 200ns. However, the PTP-based standard Ethernet network is not able to transmit packets deterministically. This is mainly due to the fact that while all nodes are synchronized, their packets have to be treated as BE (Best Effort) traffic. Consequently, this causes collisions and latencies when traffic intensifies. Therefore, as shown in Fig. 6, the proposed fully synchronized hybrid wired/wireless architecture is shown to be capable of achieving high reliability and low jitter, which are essential for synchrophasor communications.

## V. CONCLUSION

In this article, we presented a hybrid wired/wireless deterministic network architecture that integrates the TTE wired backbone network with PTP aided 5G URLLC wireless sub-networks. Specifically, high speed TTE is considered to achieve high bandwidth deterministic communication, while 5G-URLLC is employed to obtain near deterministic wireless communication access in order to remotely control PMUs and other time-critical sensory devices. Thanks to the high precision time synchronization and link reservation mechanism in both TTE and PTP aided URLLC, the proposed hybrid network is shown to be capable of supporting time-critical services with low latency and very low jitter. This guarantees reliability for Industry 4.0 as demonstrated by the robust performance of the proposed hybrid network.

ACKNOWLEDGMENT

The authors would like to thank Dr. Ya-Shian Li-Baboud for her constructive comments.